\documentclass[amsmat,amssymb,amsfonts,aps,prl,twocolumn,showpacs]{revtex4}

\usepackage{graphicx}
\usepackage{dcolumn}
\usepackage{bm}
\begin{document}

\title{ Spontaneous persistent currents in   magnetically ordered graphene ribbons}

\author{ D. Soriano (1,2), J. Fern\'andez-Rossier (1)}
\affiliation{(1) Departamento de F{\`i}sica Aplicada, Universidad de Alicante, San
Vicente del Raspeig, Spain \\ (2) Instituto de Ciencia de Materiales de Madrid,  CSIC,  Cantoblanco E28049, Madrid, Spain}

\date{\today} 

\begin{abstract} 

We present a new mechanism for dissipationless persistent charge current.
Two dimensional topological insulators  hold dissipationless spin currents in their edges so that,  for a given spin orientation, a net charge current flows which  is exactly compensated by the counter-flow of the  opposite spin.   Here we show that  ferromagnetic order in the edge upgrades the  spin currents  into persistent charge currents,  without  applied fields.
 For that matter, we study  an interacting graphene zigzag ribbon with   spin-orbit coupling.
We find three electronic phases with magnetic edges  that carry currents  reaching  $0.4$nA,  comparable to persistent currents in metallic rings,  for the small  spin orbit coupling in graphene. One of the phases is a valley half-metal.

\end{abstract}

\maketitle
Ordered electronic phases can emerge in condensed matter with    properties fundamentally  different from those of the constituent atoms. 
Two main different scenarios are known  that result  in the emergence of non-trivial electronic order.  On one side,  spontaneous symmetry breaking driven by many-body  interactions   which  accounts for the existence of the crystalline order in solids and the variety of  ordered electronic phases they can present,   like  superconductivity and ferromagnetism \cite{Anderson}.  On the other side,  topological order, which accounts for the robust quantized properties of the electron gas in the Quantum Hall regimes, and, more recently,  on the properties of the   so called  topological insulators\cite{reviews-TI,Kane-Mele1,Kane-Mele2,TI}.   Whereas Integer Quantum Hall state is driven by an external magnetic field,  topological insulators are driven by spin orbit interaction. 
They  are different from conventional insulators because of  their   conducting surface (or edge)  spin states which can be either chiral or spin-filtered and are  robust with respect to time-reversal symmetric perturbations \cite{reviews-TI, Kane-Mele1,Kane-Mele2,TI,Sinitsyn06}.  

The experimental finding of topological insulators\cite{Konig07,Chen09} and their exotic surface (edge) states,   motivates   the general questions of whether and how electronic interactions could produce electronic phase transitions in the surface or edge states of a topological insulator, and what would be the consequences of such symmetry breaking. 
Here we address these  questions in the case of a graphene   ribbon with zigzag edges, a system that atracts enormous interest
\cite{Kane-Mele1,Kane-Mele2,Nakada96,Fujita96,BreyFertig,Son06,Cohen06,Pablo07,RMP07,Valley,Gun07,Yaz07,Dai08,JFR08,kim08,Fede09} in the context of spintronics\cite{Kane-Mele1,Kane-Mele2},  magnetoelectronics\cite{Fujita96,Son06,Cohen06,Gun07,Yaz07,JFR08,Fede09} and valleytronics\cite{Valley} . 
When described with a one orbital tight-binding model, graphene zigzag ribbons are conducting because of
two  degenerate  almost flat bands associated to states localized at the edges \cite{Nakada96}. Importantly, the zigzag states preserve the valley character \cite{BreyFertig,Valley} of  two dimensional graphene.  When considered  separately, the  effect of electronic repulsion and  spin orbit interactions over the system is dramatic and has been widely studied.  On one side, it was soon recognized that  electron Coulomb repulsion, in the Hubbard model,   results in ferromagnetic order  at the edges of the ribbon   \cite{Fujita96}.   In the ground state the edges are counter-polarized and according to both density functional calculations \cite{Son06} and to mean field Hubbard model\cite{JFR08}, a gap opens so that the ribbon is an antiferromagnetic insulator. 

When Coulomb repulsion is neglected,   spin orbit (SO) interaction  opens a gap in the spectrum of bulk graphene\cite{Min06,Yao07} which is accompained by the emergence of  spin filtered edge states at the Fermi energy\cite{Kane-Mele1,Kane-Mele2}.  These two features, a spin-orbit driven gap and the emergence of topologically robust edge states, are the  main properties of   Topological  Insulators\cite{reviews-TI,Kane-Mele1,Kane-Mele2,TI,Sinitsyn06}.   
Thus, the Coulomb  driven and the SO driven phase have very different magnetic and conducting properties.  
Here  we address  the  electronic properties of the ribbon when both Coulomb repulsion and spin orbit coupling are considered within a mean field Hubbard model\cite{Fujita96,JFR08}  with Kane-Mele spin orbit coupling\cite{Kane-Mele1,Kane-Mele2}. We report  two  main findings.  First, in the presence of SO coupling the states with counter-polarized ferromagnetic edges  break valley symmetry
and, above a critical  SO strength,   the gap closes in one valley only, resulting in a  valley half-metal.
  Second, in the presence of SO coupling ferromagnetic edges give rise to  charge currents in the edges without an applied magnetic field.   

We consider the one orbital  Hubbard model in a honeycomb  zigzag ribbon, at half filling, with the addition of the Kane-Mele SO coupling\cite{Kane-Mele1}:
\begin{equation}
{\cal H}=  \sum_{i,j,\sigma} t_{i,j}(\sigma) c^{\dagger}_{i,\sigma}c_{j,\sigma} 
+ U \sum_{i} n_{i\uparrow} n_{i\downarrow}
\label{HKM}
\end{equation}

The hopping matrix  $t_{ij}(\sigma)$ in the first term accounts  both for the standard first-neighbor spin-independent hopping $t$ and the
Kane-Mele-Haldane (KMH) second neighbor spin-orbit coupling\cite{Kane-Mele1,Kane-Mele2,Haldane88}. The amplitude of the latter is given by
 the expression $i t_{KM} \sigma\hat{z}\cdot\left( \vec{d}_1\times\vec{d}_2\right)$, where $\vec{d}_{1,2}$ are unit vectors along the direction of the bond that connect atom $i$ and $j$ with their common first neighbor (see figure 1a), $\hat{z}$ is the unit vector normal to the ribbon plane,  and $\sigma=\pm 1$ indexes the spin projection along $\hat{z}$.
  For a flat ribbon, the SO term conmutes with $\sigma_z$.   The  second term in eq. (\ref{HKM}) describes Coulomb repulsion between electrons in the Hubbard approximation.  We treat it in a mean field approximation  so that we end up with an effective single particle Hamiltonian where the electrons interact with a spin dependent potential that is calculated self-consistently, $ U\sum_i \left( n_{i,\uparrow}\langle n_{i,\downarrow}\rangle +  n_{i,\downarrow}\langle n_{i,\uparrow}\rangle\right) $, so that $\sigma_z$ is a good quantum number.   Zigzag  ribbons are defined by  $N_y$, the number of zigzag chains which yield a total of $2 N_y$ atoms per unit cell in a one dimensional crystal (fig. 1a). Importantly, the top and bottom edges belong to the two different triangular sub-lattices that define  the honeycomb lattice.  For a given wavenumber $k$ and spin $\sigma$   the mean field Hamiltonian has $2 N_y$ states $\Psi_{k\sigma\nu}(y)$  with energy $\epsilon_{\sigma \nu}(k)$.

\begin{figure}
[t]
\includegraphics[width=0.90\linewidth,angle=0]{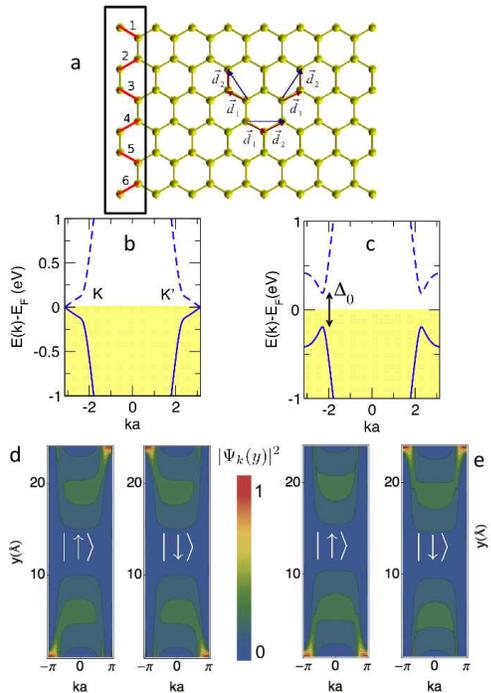}
\caption{ \label{fig1} 
 (a) Honeycomb zigzag ribon with $N_y=6$ atom rows. The   second neighbour hopping vectors $\vec{d}_i$ are shown.  
   (b) Low energy  bands for the Spin Hall phase in  $N_y=12$ ribbon ($U=0,t_{KM}=0.01$ eV). (c)  Low energy bands for the  same ribbon with counter-polarized ferromagnetic edges ( $U=3$ eV and $t_{KM}=0$). (d) Contour map of the valence band wave function $|\Psi_{k\sigma\nu}(y)|^2$ for spin up (left panel) and spin down (right panel).  (e) Same than (d) for the magnetic ribbon. }
\end{figure}

We consider first  the model in the two limit cases, $U=0, t_{KM}>0$ and $t_{KM}=0, U>0$  for a ribbon with $N_Y=12$.   The highest occupied and lowest empty  energy bands of the  $U=0, t_{KM}>0$ case are shown in figure 1b. The special spin filtered edge states are the linear bands crossing the Fermi energy ($E_F$) The wave function squared, $|\Psi_{k,\sigma,\nu}(y)|^2$, of the valence band states  are represented in figure 1d for $\sigma=\uparrow$ and $\sigma=\downarrow$  respectively.  It is apparent that $\sigma=\uparrow$  ($\sigma=\downarrow$) electrons can only be in the top (bottom) edge for positive (negative) velocity  states. Spin $\uparrow$ and $\downarrow$ electrons can also be in the bottom edge, but with velocity opposite to that of the top edge.  Thus, the ribbon is conducting, the edges are not spin-polarized, but the states at the Fermi energy carry a net spin current.  The electronic structure of the same ribbon, but now taking $t_{KM}=0, U>0$, is radically different. Instead of the linear spin-filtered bands there is a gap, $\Delta_0$  so that the system is insulating. 
 The wave-functions of the valence band are shown in figure 1e. 
They do not show  correlation between spin and velocity, but they are edge-sensitive: top (bottom) edge is ferromagnetic with spin up (down) majority. The solution with reversed spins is equally valid. 

In the light of figure 1,  the question of how these competing electronic phases merge when both the Coulomb repulsion and the SO coupling are present calls for an answer. In figure 2 we show the energy bands for  three cases, all  with $U=t=3$eV and $N_y=12$. 
In the upper panels we show two cases with magnetic edges with total spin zero, one with with $t_{KM}=0.01 t$ and the second with $t_{KM}=0.03 t$.   As shown in the inset, the magnetic moments are localized in the edges. 
 In the small $t_{KM}$ case, the system in an insulator, but the inter-edge gaps $\Delta_0$  are now valley dependent. As the SO coupling increases the gap in one of the valleys closes completely, yet the edges are magnetic. This phase is radically different from the SO free case:  a valley half-metal with magnetic edges antiferromagnetically oriented. In figure 2d we show how the magnetic moment in the edge atoms, $m=\frac{\langle n_{\uparrow}\rangle-\langle n_{\downarrow}\rangle}{2}$ is depleted as the strength of the spin orbit coupling is increased, reflecting the competition between the two terms in the Hamiltonian.  Above a certain value of $t_{KM}$, the magnetic moment vanishes altogether and a Spin Hall insulator phase with conducting edge states  identical to that with $U=0$ is obtained.   The calculated phase diagram in the $(t_{KM},U)$ plane showing the three different phases with zero total spin is shown in figure 3a, for the case of $N_Y=12$.    In figure 2c we show the conduction and valence band of the ferromagnetic phase with copolarized edges (see inset of fig. 2c).  These bands preserve valley symmetry and intersect the Fermi energy, so that the edge states are conducting. 
 
 \begin{figure}
[hbt]
\includegraphics[width=0.9\linewidth,angle=0]{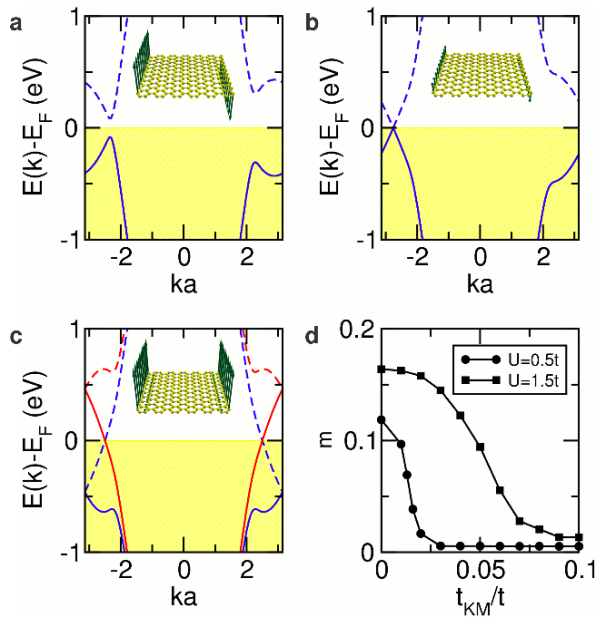}
\caption{ \label{fig2} (Color Online).  
 Valence and conduction band of a $N_y=12$ ribbon with $U=t=3 eV$  for three cases: (a) AF insulator with  $t_{KM}=0.01t$, (b), AF valley half metal $t_{KM}=0.03t$ and (c) Co-polarized ferromagnetic edges  $t_{KM}=0.03t$.  Insets:calculated magnetic density along the ribbon cell. (d) Depletion of the edge magnetic moment as the SO coupling $t_{KM}$ is increased, for two different values of $U/t$.}
\end{figure}

The evolution of the valley symmetry breaking, reflected by the different size of $\Delta_0(K)$ and $\Delta_0(K')$,  is shown in figure 3b. 
 The valley symmetry breaking  can be understood using perturbative arguments. Close to the Dirac point
the  Kane-Mele Hamiltonian can be approximated\cite{Kane-Mele1} by  $V_{KM}=3\sqrt{3}t_{KM} \sigma_z  \tau _z \lambda_z$, where $\sigma_z$, $\tau_z$ and $\lambda_z$ are the spin, valley and sub-lattice index respectively.  
 Let us consider first the $(t_{KM}=0,  U>0)$ antiferromagnetic phase (figure 1c,e) as starting point.  The valence band is made of states with $\sigma_z=\uparrow$ and sublattice $ \lambda_z=+$  (bottom edge) and states with $\sigma_z=\downarrow$ and sublattice $ \lambda_z=-$ (upper edge), so that in both cases the product $\sigma_z   \lambda_z$ has the same sign. Now it is apparent that the expectation value of $ \sigma_z  \tau _z \lambda_z$ has opposite signs in opposite valleys, so that in one case the gap opens and in the other closes, as shown  in figure 2a. 
 %

\begin{figure}
[hbt]
\includegraphics[width=0.9\linewidth,angle=0]{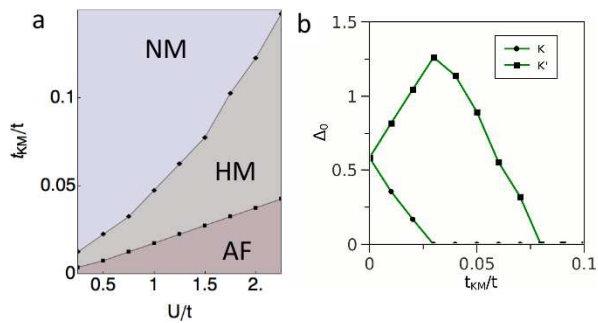}
\caption{ \label{fig3} (Color online). (a) Phase diagram  for the  AF insulating,
 AF valley half-metal (HM) and  non-magnetic (NM) phases for the $N_y=12$ ribbon.  (b) Evolution of the gaps $\Delta_0(K)$ and $\Delta(K')$  as  a function of the SO coupling.}
\end{figure}

In the  non-interacting Kane-Mele model a non-equilibrium current  $I$  induces spin accumulation $m$ in the edges\cite{Kane-Mele1}.
This can be quantified as follows.  We assume that a population of non-equilibrium extra electrons occupies the positive velocity states, so that the Fermi energy is increased by  $\delta E_F=eV$.  Using Landauer formula we have  $I=2\frac{e}{h}\delta E_F$, half of which goes on the top edge.  
  From the  non-interacting conduction band  dispersion  $\epsilon_k  = \hbar v_F k $,
 we obtain $\delta k_F$ and the corresponding  change in density due to $\delta E_F$,  $\delta n= \frac{\delta q}{a}= \frac{1}{\pi} \delta k_F$.
 Since $\delta E_F$ is  small, this extra density goes as spin $\uparrow$  to one edge and spin $\downarrow$  to the other. Thus, half of the extra charge goes to each edge  fully spin-polarized, so that the edge magnetic moment reads $\delta m=\frac{1}{4}\delta q=\frac{1}{4\pi} \delta k_Fa$. We can write the  current in a given edge as
 \begin{equation}
I_{edge}=4\pi \frac{e v_F}{a}m
\label{Ianal}
\end{equation}

We now show that this picture survives in the interacting case in equilibrium.  We find that  in the ground state of the the magnetically ordered topological insulator phases shown in figure 2,  charge currents flow on the edges, the total current accross a unit cell being null, in agreement with general theorems\cite{Bohm}. The current  operator
is given by the sum of link currents associated with all the sites $b$ connected to $a$  by single particle hopping:
\begin{equation}
\hat{I}_a = \frac{ei}{\hbar} \left( \sum_{b\sigma} t_{a,b}(\sigma) c^{\dagger}_{a,\sigma}  c_{b,\sigma} - t_{b,a}(\sigma) c^{\dagger}_{b,\sigma}  c_{a,\sigma}\right)
\end{equation}
For a given eigenstate $\Psi_{k\sigma\nu}$ of the mean field Hamiltonian the current across the link $ab$ reads
\begin{equation}
I_{ab}\left[\Psi\right] =
2 \frac{e }{\hbar N}Im \left(\Psi_{k\sigma\nu}^*(b)  \Psi_{k,\sigma,\nu}(a)    t_{ab}(\sigma) e^{i \phi_{ab}} \right)
\end{equation}
where $\phi_{ab}=0$ for $a$ and $b$ in the same cell,  $\phi_{ab}=\pm ka$ if $a$ and $b$ are  in adjacent cells, and $N$ is the number of cells in the crystal. 
The expectation value of this operator in the many-body ground state is obtained summing over all occupied bands:
$\langle I_{ab}\rangle =\sum_{k,\nu,\sigma} f\left(\epsilon_{\nu\sigma}(k)\right) I_{ab}$.
The  average current   is   defined in the links of any pair of atoms connected by hopping in the one-body hamiltonian. 
In figure 4 we plot the ground state  current map  for the $N_y=12$ ribbon for the AF insulating phase (fig 2a) and the ferromagnetic conducting phase (fig. 2c). The AF valley-half metal (not shown) is very similar to the AF insulator.  The three electronic phases described in figure 2 present edge current of similar magnitude. Whereas in the FM case,  currents flows in opposite directions in opposite edges, in the AF phases current runs parallel in the two edges.
These results can be rationalized as if the magnetization plays the role of an external magnetic field. Thus,  in the ferromagnetic case  current flows is the same than in a Quantum Hall bar.  In the antiferromagnetic phases, though, current flows parallel in the two edges, as if a magnetic field was pointing along opposite directions in the two edges.

\begin{figure}
[t]
\includegraphics[width=0.7\linewidth,angle=0]{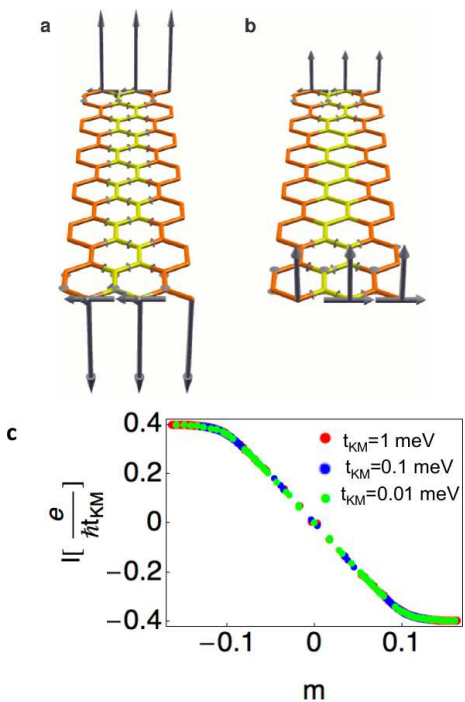}
\caption{ \label{fig4} Current and magnetization  maps for  (a) the AF insulating $N_y=12$ ribbon and   (b) the FM conducting one, both with $t_{KM}=0.03t$ and $U=1.0t$. The current between two atoms $\vec{r}_a$ and $\vec{r}_b$ is plotted as a vector along the line 
$\vec{r}_b-\vec{r}_a$, starting in the midpoint.   (c), Edge charge current, in units of $\frac{e}{\hbar} t_{KM}$,  as a function of the magnetic moment of the edge, for 3 values of $t_{KM}=10^{-2}$, $10^{-1}$ and $1 meV$, for the AF phases.}

\end{figure}

The magnitude of the top-edge current,  normalized by $\frac{e}{\hbar} t_{KM}$ , as a function of the top edge magnetization $m$   collapses for several values of $t_{KM}$ (figure 4c).   For small $m$ the curve is linear, $I_{edge}\simeq -4  \frac{e}{\hbar} t_{KM}m $, in qualitative agreement with the analytical result of eq. (\ref{Ianal}), since  $v_F=\gamma t_{KM} a/\hbar$  with  $\gamma \simeq 6.5$.  At larger $m$ the edge current saturates to  $|I_{edge}|\simeq 0.4  \frac{e}{\hbar} t_{KM}$.  If we take $t_{KM}=10 \mu eV$, close to the small values obtained by ab-initio calculations  \cite{Gmitra09}, we obtain   an edge current of $\simeq0.4 nA$, well within reach of state of the art persistent current detection \cite{Bles09}.

In conclusion, we propose a new mechanism for persistent charge currents. It involves the edge states of a two dimensional topological insulator with spontaneous ferromagnetic order induced by Coulomb interactions. We propose that this scenario occurs naturally in graphene zigzag ribbons. We find three new electronic phases in that system that   combine ferromagnetic order  and spontaneous charge current flow, both localized in the zigzag edges.   They arise from the interplay of Coulomb repulsion and spin orbit coupling. When the ferromagnetic edges are counter-polarized, the valley symmetry is broken and,  above  a critical strength of the spin orbit coupling, the system goes from an insulating to a valley-half-metal phase. 
In the three phases,  current flows as if there was a real magnetic field perpendicular to the sample along the direction of the magnetization edge.   The question of weather our findings can be generalized to magnetically doped topological insulators will be addressed in future work.


This work has been financially supported by MEC-Spain (Grant Nos.
MAT07-67845 and CONSOLIDER CSD2007-0010). We are indebted to J. J.  Palacios,  D. Gosalbez Martinez,  C. Untiedt, F. Guinea, L. Brey and A. S. Nu\~nez, for fruitful discussions.

\end{document}